# Point Transformations and Relationships Among Linear Anomalous Diffusion, Normal Diffusion and the Central Limit Theorem

Donald Kouri[1,2,*], Nikhil Pandya[1,2], Cameron L. Williams[2], Bernhard G. Bodmann[2], Jie Yao[3]

[1] Department of Physics, University of Houston, Houston, Texas, USA
[2] Department of Mathematics, University of Houston, Houston, Texas, USA
[3] Department of Mechanical Engineering, Texas Tech University, Lubbock, Texas, USA

## ABSTRACT

We present new connections among linear anomalous diffusion (AD), normal diffusion (ND) and the Central Limit Theorem (CLT). This is done by defining a point transformation to a new position variable, which we postulate to be Cartesian, motivated by considerations from super-symmetric quantum mechanics. Canonically quantizing in the new position and momentum variables according to Dirac gives rise to generalized negative semi-definite and self-adjoint Laplacian operators. These lead to new generalized Fourier transformations and associated probability distributions, which are form invariant under the corresponding transform. The new Laplacians also lead us to generalized diffusion equations, which imply a connection to the CLT. We show that the derived diffusion equations capture all of the Fractal and Non-Fractal Anomalous Diffusion equations of O'Shaughnessy and Procaccia. However, we also obtain new equations that cannot (so far as we can tell) be expressed as examples of the O'Shaughnessy and Procaccia equations. The results show, in part, that experimentally measuring the diffusion scaling law can determine the point transformation (for monomial point transformations). We also show that AD in the original, physical position is actually ND when viewed in terms of displacements in an appropriately transformed position variable. We illustrate the ideas both analytically and with a detailed computational example for a non-trivial choice of point transformation. Finally, we summarize our results.

## 1  Introduction

The Central Limit Theorem (CLT) is closely related to the "normal diffusion" (ND) process and is relevant to many random processes (GILLESPIE 2013, FELLER 1971, HARRISON 1971, TSALLIS 2005, PLASTINO 2011). The characteristic of ND is the fact that the mean square displacement (MSD) of a diffusing particle scales with time as $t^1$. The CLT together with the assumed zero mean and independence of increments implies that regardless of the microscopic model, the transition probability at large scales is well approximated by a Gaussian (EINSTEIN 1905). When this Gaussian kernel is convolved with an initial probability distribution, it evolves under time and after scaling tends to a Gaussian. We associate the evolution of the initial probability distribution in time with the semigroup generated by the Laplacian, $\frac{\partial^2}{\partial x^2}$ (GOLDSTEIN 1957). Because of the spectral properties of the standard Laplacian, the time dependent (Gaussian) solution is an "attractor" solution to the ND equation. However, in recent years, great interest has been focused on the phenomena of "anomalous diffusion" (AD) (PLASTINO 2011, TSALLIS 2009, METZLER 2000, 2004, 2014, BEN-AVRAHAM 2000, SORNETTE 2001, RISKIN 1984, O'SHAUGHNESSY 1985, PLEROU 2000, BARKAI 2014, WANG 1992, KÄRGER 1988, GREBENKOV 2007, GEFEN 1983, ZANETTE 1995, BOHR 1993, KLEUKE 2014, KLAFTER 2005, SANCHO 2009, LAPAS 2008, LIU 2008, HE 2008, AMIR 2010, SAGI 2012, HANSEN 2013, AGRAWAL 2015, CISTERNAS 2016, GOLESTANIAN 2009). As can be seen from the above references, such processes are observed for a host of physical, chemical and biological phenomena. The AD processes of interest to us here are those for systems with an "effective position dependent diffusion coefficient", resulting in time evolution governed by a generalized Laplacian (e.g., TSALLIS 2005, PLASTINO 2011, O'SHAUGHNESSY 1985). Herein, we provide the derivation of a general, infinite family of such Laplacians, including new ones not included in the landmark study of O'Shaughnessy and Procaccia (O'SHAUGHNESSY 1985). In the case of AD, the characteristic feature is the fact that the MSD scales with time as $t^{1/\beta}$, where β is a real number. If $0.25 \leq \beta < 1$, the process is termed "super diffusion" and if 1 < β, the process is termed "sub-diffusion". If $0 < \beta < 0.25$, the diffusion is termed ballistic. O'Shaughnessy and Procaccia were the first to show that many AD systems have exact, attractor solutions, called "stretched Gaussians": $\rho \propto \exp[-x^{2+\vartheta}/4Dt]$ (O'SHAUGHNESSY 1985). Here, D is a "constant effective diffusion coefficient". We postulate that the fact that these are computationally observed to be attractor solutions of the relevant differential equations implies that the standard CLT is "hidden" and applies also to AD associated with the diffusion equations we derive. In this paper, we show that this is indeed true.

The paper is organized as follows. In Section II, we summarize the solution of the ND equation, emphasizing its dependence on the Fourier transform (FT) of the probability function (the characteristic function) and noting that the characteristic function plays a key role in the proof of the CLT. In Section III, we discuss important connections among super symmetric quantum mechanics (SUSY), Heisenberg's uncertainty principle (HUP), point transformations (PT) (in order to deduce generalizations of the FT and the related Laplacians), and the CLT for diffusion in the new canonically conjugate "position". In Section IV, we explore the relations among AD, ND, scaling laws and PT's. In Section V, we provide a computational example for the point transformation, $W(x) = x + x^3$, that illustrates the relation of the time dependence of the MSD to the functional form of the PT,

W(x). Section VI discusses the detailed relationship between our diffusion equations and those in (O'SHAUGHNESSY 1985). Finally, in Section VII, we summarize our results.

## 2 The Normal Diffusion Equation, Laplacian Semigroup and the Central Limit Theorem

In the case of ND as a random process, the proof of the CLT is most readily based on the characteristic function, i.e., the FT of the probability distribution (KLEUKE 2014). At its heart is the fact that the Gaussian is invariant under the FT. This is interesting because the ND equation is also exactly solved using the FT of the diffusion equation. This makes use of the fact that the Laplacian satisfies

$$\frac{d^2}{dx^2} e^{\pm ikx} = -k^2 e^{\pm ikx} , \tag{1}$$

where $e^{ikx}$ is the x-representation of the momentum eigenket, $k^2 \geq 0$ and $e^{-ikx}$ is the k-representation of the position eigenket. The FT kernel is therefore also an improper eigenstate of the standard Laplacian. This spectral relation ensures that the Laplacian generates a semigroup. It results in the "directional" restriction of solutions to the ND equation: the solution is stable only for increasing time. The semigroup does not possess a bounded inverse operator. It is also related to the attractor character of the solution to the ND equation (GOLDSTEIN 1957). The Gaussian minimizes the HUP for position and momentum and this provides a rigorous basis for the FT (WILLIAMS 2017a, KOURI 2017). The diffusion equation,

$$\frac{\partial}{\partial t} \rho(x,t) = D \frac{\partial^2}{\partial x^2} \rho(x,t) , \tag{2}$$

where D is constant, becomes

$$\frac{\partial}{\partial t} \hat{\rho}(k,t) = -k^2 D \, \hat{\rho}(k,t), \tag{3}$$

under the FT $\left( \hat{f}(k) = \frac{1}{\sqrt{2\pi}} \int_{-\infty}^{\infty} dx \exp(-ikx) f(x) \right)$. This is easily integrated (in time), yielding

$$\hat{\rho}(k,t) = \exp(-Dk^2 t) \hat{\rho}(k,0) . \tag{4}$$

The inverse FT of a product of functions of k yields the exact convolution solution

$$\rho(x,t) = [4\pi Dt]^{-1/2} \int_{-\infty}^{\infty} dx' \exp[-(x-x')^2/4Dt] \rho(x',0) . \tag{5}$$

Clearly, if $\hat{\rho}(k,0)$ is equal to 1, $\rho(x',0) = \delta(x'-x_0)$ (the diffusing particle is localized initially at $x = x_0$, with time dependence given by $\rho(x,t) = C \exp[-(x-x_0)^2/4Dt]$). For any initial probability distribution whose FT is differentiable, the long-time behavior will be dominated by the overall Gaussian envelope $\exp(-x^2/4Dt)$ (simply expand the square in the exponent in Eq. (5) and factor out $\exp(-x^2/4Dt)$). The $t^{-1/2}$ in the normalization arises from requiring that the probability distribution be normalized to 1 under the measure dx. Then the MSD varies as $2Dt$ (note, the slope of the MSD is not necessarily equal to one). We thus observe the role of the CLT as implying that the time dependence of any initial distribution, evolving according to Eq. (2), will tend to a Gaussian distribution envelope after a sufficiently long time. We stress that the characteristic function proof of the CLT also rests on the facts that the Gaussian is invariant under the FT and the FT satisfies the convolution theorem. These properties are related to the semigroup structure of the standard Laplacian.

## 3   Super Symmetric Quantum Mechanics, Heisenberg's Uncertainty Principle, Point Transformations, Canonical Quantization, Generalized Fourier Transforms and Laplacians

Recently, we have explored connections among super symmetric quantum mechanics (SUSY), Heisenberg's uncertainty principle (HUP), point transformations (PT) and generalized Fourier transforms (GFT) (DIRAC 1958, WILLIAMS 2017a, KOURI 2017). We briefly summarize the relevant details here. A mathematically rigorous derivation of the FT (and its generalization) from the HUP has recently been given (WILLIAMS 2017a, KOURI 2017). The fundamental starting point is that the minimum uncertainty state, $|\psi_{min}\rangle$, satisfies

$$\hat{x}|\psi_{min}\rangle = -i\hat{k}|\psi_{min}\rangle . \quad (6)$$

The state $|\psi_{min}\rangle$ cannot be an eigenstate simultaneously of both the position and momentum operators since they do not commute ($[\hat{x},\hat{k}] = i\hat{1}$). Explicitly, we note that Eq. (6) reminds one of the property of simultaneous eigenvectors of commuting operator observables, $\hat{A}, \hat{B}$:

$$\hat{A}|a,b\rangle = a|a,b\rangle , \quad (7)$$

$$\hat{B}|a,b\rangle = b|a,b\rangle , \quad (8)$$

$$\Rightarrow \quad \hat{A}|a,b\rangle = \frac{a}{b}\hat{B}|a,b\rangle . \quad (9)$$

*The uncertainty product for such operators as $\hat{A}, \hat{B}$ is, of course, the absolute minimum value of zero and the simultaneous eigenvector is the only vector appearing in the above equations. The fact that the uncertainty product for position and momentum is positive definite results from the fact that neither $\hat{x}|\psi_{min}\rangle$ nor $-i\hat{k}|\psi_{min}\rangle$ can be simply proportional to the state*

*$|\psi_{min}\rangle$. In fact, the uncertainty minimizing state is orthogonal to the state produced by the action of the position and momentum operators (unless the average values of these operators is non-zero. Then the minimizing state is orthogonal to both $(\hat{x}-\langle x\rangle)|\psi_{min}\rangle$ and $(\hat{k}-\langle k\rangle)|\psi_{min}\rangle$! The best one can achieve for the two non-commuting operators is that their actions on the minimizing vector produce the same abstract vector, different from the minimizing state, and the new vector produced must have the same functional form in either the x- or k-representations. This immediately implies the form invariance of the Gaussian under the FT. In Appendix A, we show that this is a general property of the minimizing state vector for any pair of non-commuting, canonically conjugate Cartesian-like operators (JUNKER 1996, KLAUDER 2015).*

The 1-D SUSY formalism is built on a program to make all systems (on the domain -∞ < x < ∞) look as much as possible like the harmonic oscillator, whose ground state also satisfies Eq. (6). In the ladder operator notation, this equation is

$$\frac{1}{\sqrt{2}}[\hat{x}+i\hat{k}]|\psi_{min}\rangle = \hat{a}_{HO}|\psi_{min}\rangle = 0 \Rightarrow |\psi_{min}\rangle = |\psi_0\rangle \tag{10}$$

In the coordinate representation, $\hat{a}_{HO}$ is $(x+d/dx)/\sqrt{2}$.

In 1-D SUSY, the ground state is restricted to the form

$$\psi_0(x) = \psi_0(0)\exp[-\int_0^x dx' W(x')] \tag{11}$$

for W's such that the ground state is $L^2$. Choosing the zero of energy to be that of the ground state, one finds that the potential of the system is given by

$$V = [W^2 - \frac{dW}{dx}]/2, \tag{12}$$

and in terms of the position representation, the ground state satisfies

$$\frac{1}{\sqrt{2}}[W(x)+d/dx]\psi_0 = \hat{a}_W\psi_0 = 0. \tag{13}$$

As noted previously, this immediately implies that the ground state, $\psi_0$, minimizes the uncertainty product $\Delta W \Delta k$ ($\hbar=1$) and suggests that one ought to be able to obtain a new transform under which $\psi_0$ will be form invariant (WILLIAMS 2017a, KOURI 2017, CHOU 2012). Unfortunately, since the commutator of W and $d/dx$ is proportional to $dW/dx$, which is constant only for W = x + constant, the desired transform kernel cannot be the simple FT kernel. As a consequence, we do not actually use the SUSY expressions directly to derive the desired transform. We remark that in the SUSY community, W is interpreted as a "super potential", based on Eq. (9). In our approach, we shall interpret W to be a generalized "position" variable, based on the facts that: (1) $\psi_0$ minimizes $\Delta W \Delta k$ and (2) the quintessential choice of W is that for the harmonic oscillator, where W is precisely the particle's "generalized position" (CHOU 2012, KLAUDER 2015, WILLIAMS 2017a, 2017b, 2018).

Recently, in KOURI (2017), we explored choosing W to represent a point transformation of the usual position, such that (1) the domain of W is $-\infty < W < \infty$, (extension to the half line can be done) (2) the domain of the canonically conjugate momentum is also $-\infty < P_W < \infty$, (3) the transformation is invertible and (4) both W and $P_W$ can be interpreted as Cartesian-like coordinates. One expression that satisfies the above is the polynomial KOURI (2017)

$$W(x) = \sum_{j=1}^{2J+1} a_j x^j \quad , \tag{14}$$

in which all coefficients are positive semi-definite and each even-power coefficient is always less than the next higher odd power coefficient. This results in a monotonic increasing behavior that is invertible almost everywhere *(if one requires that the coefficient of the first power of x is positive definite, then the expression is invertible everywhere)*.

Assuming that W(x) satisfies the above conditions KOURI (2017), the classical canonically conjugate momentum, $P_W$, is required to satisfy

$$\{W, P_W\} = 1 \quad , \tag{15}$$

where $\{,\}$ is the Poisson bracket with respect to the original Cartesian-like position and momentum (DIRAC 1958, KOURI 2017). It is easily shown that the classical canonical momentum can be expressed in the form

$$P_W = \frac{1}{\left(\frac{dW}{dx}\right)^{1-\alpha}} p_x \frac{1}{\left(\frac{dW}{dx}\right)^{\alpha}} + g(x) \quad . \tag{16}$$

Invoking Occam's razor, we set the integration constant along a constant-x integration path, g(x), equal to zero. Since $P_W$ above satisfies Eq. (15) for all α, we then apply Dirac's canonical quantization to obtain (in the W-representation). This quantization consists of replacing the observables by operators, the Poisson bracket by the commutator of the operators and the scalar 1 by $i\hbar$ times the identity operator DIRAC (1958):

$$[\hat{W}, \hat{P}_W] = i\hat{1}, \quad \hbar = 1, \tag{17}$$

$$\hat{W} = W, \tag{18}$$

$$\hat{P}_W = -i\,d/dW . \tag{19}$$

***We stress that these operators are manifestly self adjoint in the W-representation, with the measure dW.*** In addition, the minimizer of $\Delta W \Delta P_W$ is obviously the Gaussian, $\exp(-W^2/2)$, since

$$W \psi_{min} = -d\psi_{min}/dW . \tag{20}$$

As in the case of the original position and momentum operators, $\psi_{min}$ is invariant under the "W-Fourier transform (W-FT) kernel":

$$\langle W|K\rangle = \exp(iKW)/\sqrt{2\pi}, \quad \langle K|W\rangle = \exp(-iKW)/\sqrt{2\pi}, \tag{21}$$

$$-\infty < W < \infty, \quad -\infty < K < \infty . \tag{22}$$

This is analogous to the FT in terms of x and k. This transform satisfies the convolution theorem under the measure dW or dK. We note that in the W-representation, since $\hat{P}_W$ is self adjoint, we have the four equivalent, exact expressions:

$$-\hat{P}_W^+ \hat{P}_W = -\hat{P}_W \hat{P}_W^+ = -\hat{P}_W \hat{P}_W = -\hat{P}_W^+ \hat{P}_W^+ = \frac{d^2}{dW^2}, \tag{23}$$

$$\hat{P}_W = \hat{P}_W^+, \tag{24}$$

which are all obviously self adjoint under the measure dW. ***The above relations are at the heart of our approach.*** The transformation that preserves the W-Gaussian ($\exp(-W^2/2)$), Eq. (17), satisfies

$$\frac{d^2}{dW^2}\exp(\pm iKW) = -K^2 \exp(\pm iKW) . \tag{25}$$

Again, the W-Laplacian possesses a negative semi-definite spectrum so it is evident that this transform kernel possesses all the properties of the FT, including the fact that it supports a semigroup property.

We also note that there exists a generalized W-harmonic oscillator, with ground state satisfying (WILLIAMS (2017a), KOURI (2017))

$$\hat{a}_W^+ \hat{a}_W |\psi_0\rangle = \frac{1}{2}(-\frac{d}{dW}+W)(\frac{d}{dW}+W)\exp(-W^2/2) = 0.. \qquad (26).$$

The diffusion equation describing independent random motion in W is obviously

$$\frac{\partial \rho}{\partial t} = D \frac{\partial^2 \rho}{\partial W^2}, \qquad (27)$$

where here, D is a constant diffusion coefficient. It follows that the exact solution of Eq. (27) is obtained using the W-FT kernel Eq. (21), yielding

$$\rho(W,t) = \int_{-\infty}^{\infty} dW' \exp[-(W-W')^2/4Dt]\, \rho(W',0), \qquad (28)$$

where $\rho(W', 0)$ is any proper, initial distribution.

The "characteristic function" underlying the above expression is

$$\hat{\rho}(K,t) = \exp(-DK^2 t)\, \hat{\rho}(K,0). \qquad (29)$$

If $\rho(W',0) = \delta(W'-W_0)$, then $\hat{\rho}(K,0)$ equals one and the distribution is a Gaussian centered at $W_0$. Note that in general, $\rho(W,t)$ involves a positive function of time times a Gaussian envelope, $\exp[-W^2/4Dt]$. ***The analysis of the CLT for probability distributions in W is identical to that for the normal probability distribution (KLEUKE 2014).***

We next consider the situation where we quantize Eqs. (23) - (24) in the x-representation (***we do not invoke the chain rule***). Because of the ambiguity of operator ordering, we consider the following explicit expressions for a generalized momentum operator and its adjoint, involving the parameter α:

$$\hat{P}_W = \frac{-i}{\left(dW/dx\right)^{1-\alpha}} \frac{d}{dx} \frac{1}{\left(dW/dx\right)^{\alpha}}, \qquad (30)$$

$$\hat{P}_W^+ = \frac{-i}{\left(dW/dx\right)^{\alpha}} \frac{d}{dx} \frac{1}{\left(dW/dx\right)^{1-\alpha}}. \qquad (31)$$

***But these operators are obviously not self adjoint under the measure dx (except in the case where $\alpha=1/2$)! Normally, they are discarded and replaced by their arithmetic average.*** However, we know that any self adjoint operator remains such, regardless of the representation. Of course, the point is that the above are self adjoint in the x-representation but the measure now includes the effect of the point transformation. Thus, Eq. (30) is self

adjoint under the measure $dx(dW/dx)^{1-2\alpha}$ and Eq. (31) is self adjoint under the measure $dx(dW/dx)^{2\alpha-1}$. When alpha equals ½, the measure for self adjointness reduces to simply dx.

We remark that, as shown above in Eq. (23), in the W-domain, all four Laplacians possess the W-Gaussian as the ground state solution of the W-HO. Also, all four Laplacians satisfy Eq. (25) in the W-domain. ***Thus, all four Laplacians are not only self adjoint in the x-representation but they also are all negative semi-definite operators which have an underlying semigroup structure regardless of their explicit form in the x-domain.*** Each of the four Laplacians will be self adjoint in the x-domain but only with appropriate measures. We now explicitly explore the forms of these Laplacians in the x-domain. It is convenient to group them in two pairs of two. This is because the forms

$$\Delta_1 = -\hat{P}_W^+ \hat{P}_W = \frac{1}{(\partial W/\partial x)^\alpha} \frac{d}{dx} \frac{1}{(\partial W/\partial x)^{2-2\alpha}} \frac{d}{dx} \frac{1}{(\partial W/\partial x)^\alpha} \qquad (32)$$

and

$$\Delta_2 = -\hat{P}_W \hat{P}_W^+ = \frac{1}{(\partial W/\partial x)^{1-\alpha}} \frac{d}{dx} \frac{1}{(\partial W/\partial x)^{2\alpha}} \frac{d}{dx} \frac{1}{(\partial W/\partial x)^{1-\alpha}} \qquad (33)$$

are already manifestly self adjoint under the measure dx. The other pair is

$$\Delta_3 = -\hat{P}_W \hat{P}_W = \frac{1}{(dW/dx)^{1-\alpha}} \frac{d}{dx} \frac{1}{(dW/dx)} \frac{d}{dx} \frac{1}{(dW/dx)^\alpha} \qquad (34)$$

and

$$\Delta_4 = -P_W^+ P_W^+ = \frac{1}{(dW/dx)^\alpha} \frac{d}{dx} \frac{1}{(dW/dx)} \frac{d}{dx} \frac{1}{(dW/dx)^{1-\alpha}} \qquad . \qquad (35)$$

The above two Laplacians, $-\hat{P}_W^+ \hat{P}_W^+$ and $-\hat{P}_W \hat{P}_W$, require different measures from one another as well as from the first two Laplacians (except for $\alpha = 1/2$). We stress that both are self adjoint operators in the x-domain, with the correct measure for $\Delta_3$ being $dx(dW/dx)^{1-2\alpha}$ and the correct measure for $\Delta_4$ being $dx(dW/dx)^{2\alpha-1}$. We next note that all four of these Laplacians support HO-like Hamiltonians. We define

$$\hat{H}_1 = \frac{1}{2}\left(-\frac{1}{(dW/dx)^\alpha}\frac{d}{dx}\frac{1}{(dW/dx)^{1-\alpha}} + W\right)\left(\frac{1}{(dW/dx)^{1-\alpha}}\frac{d}{dx}\frac{1}{(dW/dx)^\alpha} + W\right) \quad (36)$$

$$\hat{H}_2 = \frac{1}{2}\left(-\frac{1}{(dW/dx)^{1-\alpha}}\frac{d}{dx}\frac{1}{(dW/dx)^\alpha} + W\right)\left(\frac{1}{(dW/dx)^\alpha}\frac{d}{dx}\frac{1}{(dW/dx)^{1-\alpha}} + W\right) \quad (37)$$

$$\hat{H}_3 = \frac{1}{2}\left(-\frac{1}{(dW/dx)^{1-\alpha}}\frac{d}{dx}\frac{1}{(dW/dx)^\alpha} + W\right)\left(\frac{1}{(dW/dx)^{1-\alpha}}\frac{d}{dx}\frac{1}{(dW/dx)^\alpha} + W\right) \quad (38)$$

$$\hat{H}_4 = \frac{1}{2}\left(-\frac{1}{(dW/dx)^\alpha}\frac{d}{dx}\frac{1}{(dW/dx)^{1-\alpha}} + W\right)\left(\frac{1}{(dW/dx)^\alpha}\frac{d}{dx}\frac{1}{(dW/dx)^{1-\alpha}} + W\right). \quad (39)$$

We note that the first two Hamiltonians are self adjoint under the measure d x. The next two are self adjoint under the measures discussed above for Laplacians 3 and 4. It is easily seen that Hamiltonians 1 and 3 will have identical ground states, satisfying

$$\left[\frac{1}{(dW/dx)^{1-\alpha}}\frac{d}{dx}\frac{1}{(dW/dx)^\alpha} + W\right](dW/dx)^\alpha \exp(-W^2/2) = 0 \quad . \quad (40)$$

Hamiltonians 2 and 4 will also have identical ground states, satisfying

$$\left[\frac{1}{(dW/dx)^\alpha}\frac{d}{dx}\frac{1}{(dW/dx)^{1-\alpha}} + W\right](dW/dx)^{1-\alpha} \exp(-W^2/2) = 0 \quad . \quad (41)$$

These follow from the fact that the ground states are annihilated by the right hand operator in the respective factored Hamiltonians. ***We stress that, in the general case, none are simple Gaussians.*** Rather, the influence of the point transformation is clearly evident. Of course, if one sets $\alpha = 0$ or 1, then we see that the "stretched Gaussian" is one of the ground states. However, there is a second (bi-orthogonal partner) ground state which, in general, will be a multi-modal distribution. Because of the monotonic increasing character of the PT, $dW/dx$ is never negative and the ground state is always positive semi-definite. This is new and it reflects the fact that our formalism is closely related to the "Coupled SUSY" formalism introduced elsewhere (WILLIAMS 2017b, 2018).

The next question of importance is: What are the eigenstates of the four x-domain Laplacians? These will provide the transformations under which the various generalized HO ground states are invariant (in analogy to the FT and the Gaussian distribution). Of course, in the "W-world", there is only the W-FT, resulting from the W-Laplacian, having only the simple W-Gaussian as the solution of the diffusion and HO equations. We begin by considering the first two Laplacians. In this case, the explicit, analytical transforms have been derived in WILLIAMS (2017a) *for monomial W's and* $\alpha=0$. In that case, the first equation is

$$-\hat{P}_W^+ \hat{P}_W \Phi_K = \frac{d}{dx} \frac{1}{(dW/dx)^2} \frac{d\Phi_K}{dx} = -K^2 \Phi_K \qquad (42)$$

where $K^2 \geq 0$ (the Laplacian is negative semi-definite and self adjoint under the measure dx). The unitary transformation was derived restricting W to the monomial form $W(x)=|x|^\beta$. The result is

$$\Phi_K = \frac{1}{2} |kx|^{\beta-1/2} \left[ J_{-1+1/2\beta}(|kx|^\beta) - i \, \text{sgn}([kx]^\beta J_{1-1/2\beta}(|kx|^\beta)) \right] \qquad (43)$$

which is unitary under the measure d x. The stretched Gaussian, $\exp(-|x|^{2\beta}/2)$ is invariant under this transform. The second transform equation is

$$\frac{1}{(dW/dx)} \frac{d^2}{dx^2} \frac{1}{(dW/dx)} \tilde{\Phi}_K = -K^2 \tilde{\Phi}_K \qquad (44)$$

and the unitary transform (under the measure d x), $\tilde{\Phi}_K$, is given by

$$\tilde{\Phi}_K = \frac{1}{2} \left[ \text{sgn}(\eta)^{\beta-1} |\eta|^{(\beta-1)/2} J_{-1/2\beta}(|\eta|^\beta) + i \, \text{sgn}(\eta)^\beta |\eta|^{(\beta-1)/2} J_{1/2\beta}(|\eta|^\beta) \right] \qquad (45)$$

where $\eta = kx$.

In this case, it is the, in general, multi-modal distribution $(dW/dx)\exp(-W^2/2)$ which is invariant.

The transformations for the second two Laplacians are easily obtained for general values of $\alpha$. For the Laplacian $\Delta_3$, the result is

$$\Delta_3 \varphi_K = -K^2 \varphi_K, \; \varphi_K = (dW/dx)^\alpha \exp(\pm iKW)/\sqrt{2\pi} \quad . \qquad (46)$$

The result for the Laplacian $\Delta_4$ is given by

$$\Delta_4 \tilde{\varphi}_K = -K^2 \tilde{\varphi}_K, \; \tilde{\varphi}_K = (dW/dx)^{1-\alpha} \exp(\pm iKW)/\sqrt{2\pi} \quad . \qquad (47)$$

It is easily seen that these are bi-orthogonally related, since

$$\Delta_3^+ = \Delta_4 \quad (48)$$

and that (in Dirac notation),

$$\langle \widetilde{\varphi}_{K'} | \varphi_K \rangle = \langle \varphi_K | \widetilde{\varphi}_{K'} \rangle = \delta(K - K'), \quad (49)$$

$$\hat{1} = \int_{-\infty}^{\infty} dK \, |\varphi_K\rangle\langle\widetilde{\varphi}_{K'}| = \int_{-\infty}^{\infty} dK \, |\widetilde{\varphi}_K\rangle\langle\varphi_{K'}| \, . \quad (50)$$

Thus, the bi-orthogonal transforms serve to carry out Fourier-like transformations of the ground states of the GHOs. The ground state solution, $(dW/dx)^\alpha \exp(-W^2/2)$, when transformed using $\widetilde{\varphi}_K$ under the measure d x, will produce the stretched Gaussian. The same is true when $(dW/dx)^{1-\alpha} \exp(-W^2/2)$ is transformed by $\varphi_K$ under the measure d x. This explicitly shows how the underlying CLT for the W-domain Laplacian and W-Gaussian will be manifested in the x-representation.

## 4    Anomalous Diffusion and Normal Diffusion

We now discuss the implications of the above for diffusion. We consider the four linear diffusion equations

$$\frac{\partial \rho_j}{\partial t} = D \Delta_j \rho_j, \; j=1-4 \quad . \quad (51)$$

In the cases $j=1,2$, the Laplacians are self adjoint under dx. When we choose monomial W's with $\alpha=0$ and j = 1, transformation under $\Phi_K$ leads to

$$\frac{\partial \hat{\rho}_1}{\partial t} = -DK^2 \hat{\rho}_1 \, . \quad (52)$$

When we choose the same alpha and j = 2, transformation under $\widetilde{\Phi}_K$ again yields

$$\frac{\partial \hat{\rho}_2}{\partial t} = -DK^2 \hat{\rho}_2 \quad . \quad (53)$$

In the cases of j = 3 and 4, we take care to recall the biorthogonal character of the transforms but it again results in equations of the same exact form as above, but now for all α. *Once the time dependent equation is in the K-domain, the solution can be obtained using the original W-domain transform, Eq. (17). This transformation possesses a convolution theorem and in the W-domain, all behave in time like normal diffusion!* We conclude that all of the linear diffusion equations derived herein are subject to the CLT associated with Gaussian distributions. The above analysis holds in general for any proper choice of PT but the above analysis for the j = 1,2 cases has only been explicitly realized for the monomial case with alpha equal to zero. In the case of j = 3,4, everything applies for general point PT's, including polynomials (KOURI 2017).

For simplicity, we shall illustrate these ideas using the specific example of the j = 3 case for alpha equal to zero. The transform equation is then

$$[\frac{1}{dW/dx}\frac{d}{dx}][\frac{1}{dW/dx}\frac{d}{dx}]\exp(\pm iKW(x)) = -K^2 \exp(\pm iKW(x)). \quad (54)$$

The diffusion equation is

$$\frac{d\rho}{dt} = D[\frac{1}{dW/dx}\frac{d}{dx}][\frac{1}{dW/dx}\frac{d}{dx}]\rho . \quad (55)$$

The corresponding x-domain GFT kernel from Eq. (21) is given by

$$\varphi_K(W) = \frac{1}{\sqrt{2\pi}}\exp(\pm iW(k)W(x)) . \quad (56)$$

When applied in the form Eq. (56), the measure is $\frac{dW}{dx}dx$. We note that requiring K to have the same functional dependence on k as W(x) on x ensures the argument of the exponential is dimensionless and that the measure for integration over K in the k-domain is $dK = dk\frac{dW(k)}{dk}$. After applying the GFT and using Eq. (25), the exact solution (in the k-domain) is

$$\hat{\rho}(W(K),t) = \exp(-W(k)^2 Dt)\hat{\rho}(W(K),0). \quad (57)$$

The inverse GFT (which satisfies the convolution theorem in the variable W) then leads to the probability distribution at time t:

$$\rho(W(x),t) = \int dW(x')\exp(-[W(x)-W(x')]^2/4Dt)\rho(W(x'),0) . \quad (58)$$

*We stress that as derived, this is a probability distribution in the variable W(x), not the variable x. The measure for normalization will be $d(W(x))$, the normalization constant will be $[4\pi Dt]^{-1/2}$ and the MSD will be proportional to $t^1$. The diffusion is normal in the variable W.*

However, Eq. (58) can also be interpreted as a probability distribution in the variable x. It is a positive definite function of x and can be normalized to 1 under the measure dx:

$$\int_{-\infty}^{\infty} dx\, \rho(W(x),0) = 1. \tag{59}$$

If, e.g., $W(x)=x^\beta$, the integration will be transformed to the variable $x \to x/t^{1/2\beta}$ and the normalization will be proportional to $t^{-1/2\beta}$. The average value of x will be zero (for symmetric initial distributions) and the MSD will scale as

$$MSD \propto t^{1/\beta}. \tag{60}$$

*This is, of course, the scaling for AD: the probability distribution is interpreted as applying to the physical position variable, x, rather than to the canonical variable, $W(x)=x^\beta$. The range $0.25 < \beta < 1$ corresponds to super diffusion, the range $0 < \beta < 0.25$ corresponds to ballistic diffusion and $\beta > 1$ corresponds to sub-diffusion. However, because the Laplacian satisfies Eq. (25), the time evolution will still be such that the CLT is satisfied. The normalization has nothing to do with the fact that the time evolution is governed by the attractor, whether we express it in the W- or the x-domain. Similar scaling results apply in the general cases.*

## 5  A Computational Example for the Polynomial $W(x)=x+x^3$

We illustrate our results by the numerical example of a polynomial choice of PT:

$$W(x) = x + x^3. \tag{61}$$

We assume the initial distribution to be

$$\rho(W,0) = \exp(-1000W^2). \tag{62}$$

This is very localized at W = 0, implying that it is centered at x = 0. The diffusion equation expressed in the x-domain is given by

$$\frac{\partial \rho}{\partial t} = D \frac{1}{1+3x^2}\frac{\partial}{\partial x}\frac{1}{1+3x^2}\frac{\partial \rho}{\partial x}. \tag{63}$$

We note that in the region $|x|<<1$, W is, to a very good approximation, equal to x but as the particle diffuses to larger magnitude values of x, it quickly transitions to being dominated by the cubic x-dependence. The W-domain results are shown in Figs. (1) - (2), where it is obvious that the diffusion is normal, since Eq. (63) becomes Eq. (27).

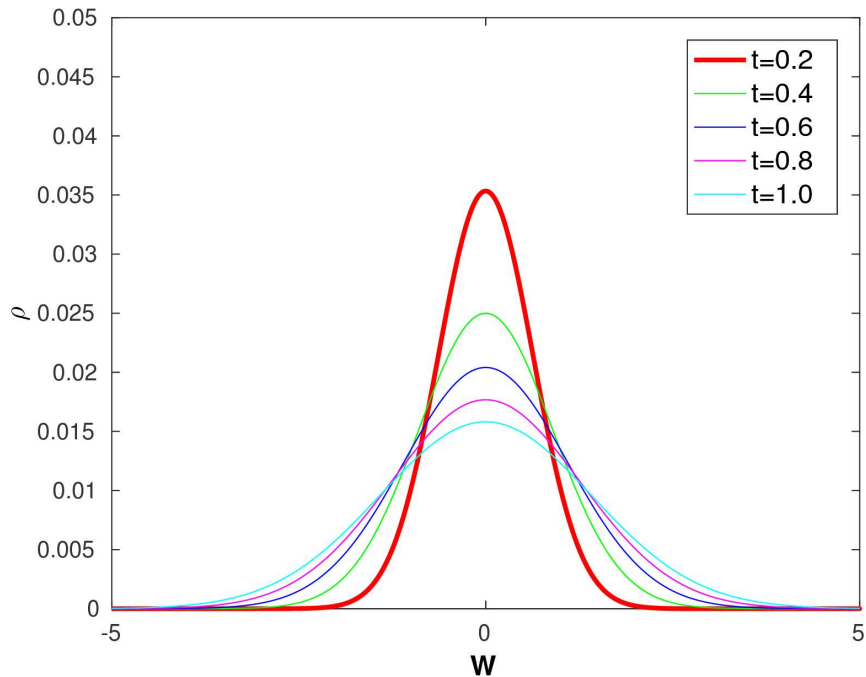

**Figure 1 Time evolution in W coordinate is normal. The different color graphs correspond to times of diffusion, as indicated in the inset.**

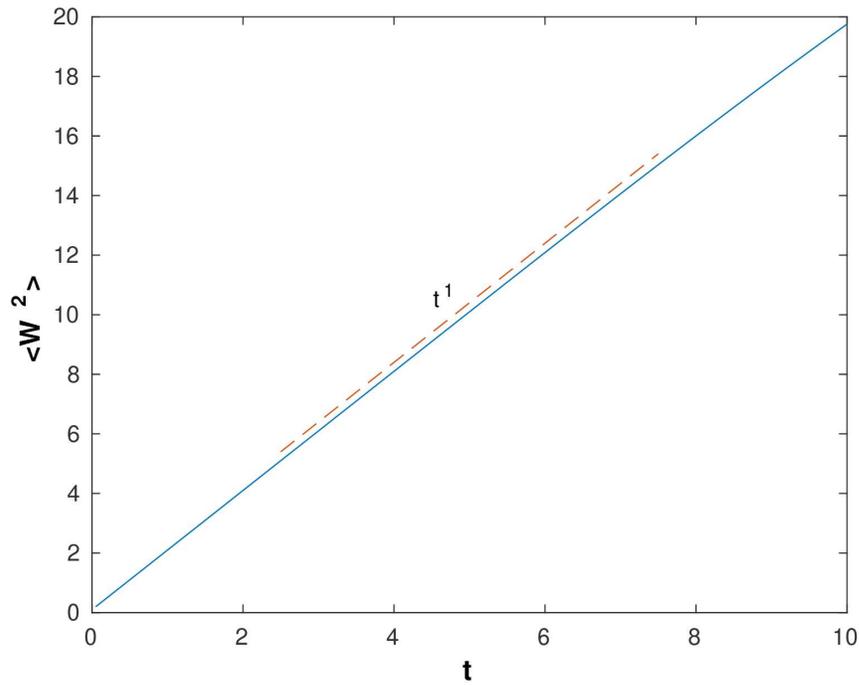

**Figure 2 MSD in W coordinate. Diffusion is normal. The solid line is the result of solving the W-domain diffusion equation. The dashed line is the curve resulting from the MSD scaling as the time, t.**

Of greater interest are the results where we interpret our density in the x-domain and use the Laplacian resulting in the diffusion equation (63). The results are displayed in Figs. (3) and (4).

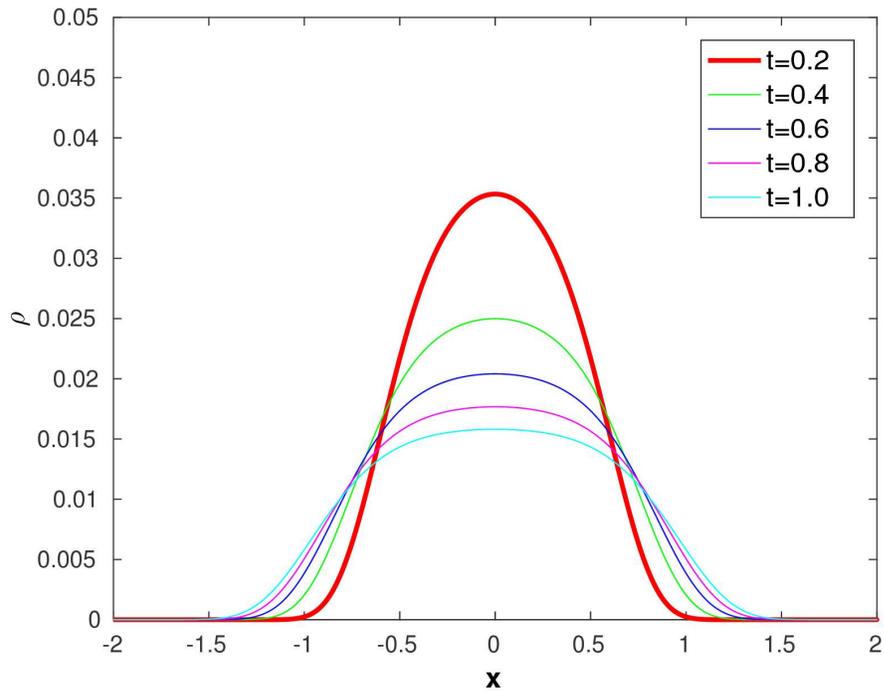

**Figure 3** Evolution in x coordinate is anomalous. The different color graphs correspond to times of diffusion, as indicated in the inset.

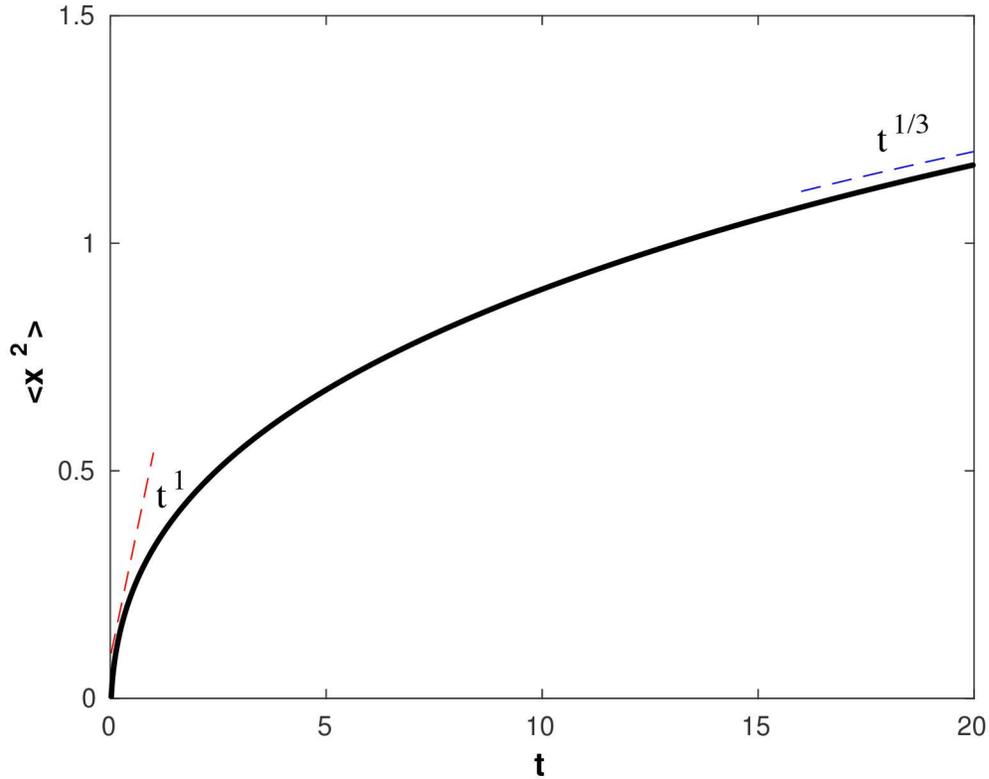

**Figure 4 Short and long time behavior of MSD in x. The early time shows ND and the later time AD. The red dashed curve beginning at t = 0 is the linear in time scaling law and the dashed blue curve at large time is the 1/3 power of t scaling law.**

It is clear that at the earliest time, the linear x-dependence dominates the Laplacian and the diffusion is essentially normal, with the normal MSD t-scaling becoming more and more accurate as t approaches zero. At longer times, the cubic x-dependence dominates the Laplacian and we see that the MSD behavior tends to that of $t^{1/3}$, just as one expects. While for more complicated polynomial PTs, such analysis will be more difficult, these results suggest that experimental measurements of the scaling as a function of time can provide insight into the appropriate PT for which the diffusion will be normal. More important, this will give information as to the best "effective" displacement variable with which to characterize and interpret the diffusion dynamics.

## 6 Relationship between Our Diffusion Equations and the O'Shaughnessy-Procaccia Equations

As formulated, our analysis and diffusion equations are very general and applicable both to fractal and non-fractal AD processes. Furthermore, for specific choices of our parameters, our equations exactly incorporate the radial diffusion equations of (O'SHAUGHNESSY 1985). This might appear somewhat strange since our equations are defined on the entire real line while those of O'Shaughnessy and Procaccia are for radial variables on the half line. In fact, it is only a subset of our equations that incorporate the O'Shaughnessy and Procaccia equations (WILLIAMS 2017a). The Laplacians are, explicitly, Eq. (32) for α = zero, were first derived on the half line and then extended to the full real axis and these are the particular equations that capture those of reference (O'SHAUGHNESSY 1985). We explicitly prove this below. The equations of interest are given by

$$\frac{\partial \rho}{\partial t} = D \frac{1}{x^{c-1}} \frac{\partial}{\partial x} x^{c-1-\vartheta} \frac{\partial \rho}{\partial x} \quad . \tag{64}$$

To transform this to our form of equations, we define the auxilliary variable

$$y = x^c , \tag{65}$$

in the equation above. It easily follows that

$$\frac{D}{x^{c-1}} \frac{\partial}{\partial x} x^{c-1-\vartheta} \frac{\partial}{\partial x} = c^2 D \frac{\partial}{\partial y} x^{2c-2-\vartheta} \frac{\partial}{\partial y} \quad . \tag{66}$$

Here, c is the dimension (non-integer c's correspond to fractal cases). If one next defines the parameter

$$\beta = \vartheta/2 - c + 2 , \tag{67}$$

then Eq. (66) becomes

$$c^2 D \frac{\partial}{\partial y} x^{2c-2-\vartheta} \frac{\partial}{\partial y} = c^2 D \frac{\partial}{\partial y} \frac{1}{y^{2\beta-2}} \frac{\partial}{\partial y} = c^2 \beta^2 D \frac{\partial}{\partial y} \frac{1}{(\partial W/\partial y)^2} \frac{\partial}{\partial y} , \tag{68}$$

where

$$W(y) = y^\beta , \quad y^\beta = x^{\beta c} \quad . \tag{69}$$

The parameters, $c, \vartheta$ remain independent of one another. Of course, this is exactly our diffusion equation for the Laplacian in Eq. (32) with $\alpha = 0$. (In addition, we note that Eq. (33) also captures the O'Shaughnessy and Procaccia equations for the specific value $\alpha = 1$.) ***It is important to note, however, that even for the simplest monomial cases, with $\alpha \neq 0, 1$, all four of our Laplacians in Eqs. (32) – (35) differ from those of reference (O'SHAUGHNESSY 1985).*** These new diffusion equations are also currently under computational study in our group.

# 7  Conclusions

***First, we note that our analysis shows that for diffusion according to any of the four diffusion equations, AD will be observed in the variable x but ND will be observed in the variable W.*** For ND to be manifested, one must analyze the results using the relevant generalized coordinate. AD as a function of x is simply "disguised ND" in W.

***Second, we have shown that when linear anomalous diffusion is analyzed in terms of the appropriate PT displacement variable, the usual CLT applies.*** Of course, this in no way alters the fact that the situation is much more complex in the case of nonlinear diffusion (TSALLIS 2005, 2009, PLASTINO 2011).

***Third, our results show that, for such AD systems, experimental determination of the anomalous scaling leads directly to the identification of the relevant generalized coordinate, W(x) (in the case of the monomial choice of W(x)).*** Certainly, for the example case of $W(x) = x + x^3$, it is also quite easy to extract the relevant PT from the numerical data. For more general polynomial PTs, it will, naturally, be more complicated but the experimental scaling should still give information as to the relevant PT and effective displacement variable.

***Fourth, we recall that the CLT is related to approximating the semigroup generated by the Laplacian operator. Because of this, we expect that any diffusion process for which a "proper" Laplacian operator exists will have an attractor solution that is invariant under the generalized transform and is an eigenfunction of the corresponding HO.*** It will be of interest to explore treatment of Levy processes using fractional powers of the generalized Laplacians.

***Fifth, we have obtained linear diffusion equations that not only capture all those of reference O'SHAUGHNESSY (1985), but include infinite families of new equations for the values*** $0 < \alpha < 1$. It will be of interest to explore whether any experimental data corresponds to these new linear diffusion equations.

***Sixth, in our analysis, we considered diffusion in a 1D "Cartesian system" with a constant diffusion coefficient, D. It readily generalizes to any number of Cartesian random variables by simply summing the generalized Laplacians for each degree of freedom. We also point out that the scaling need not be the same in each degree of freedom. This will be important for anisotropic diffusion processes.*** It is also possible to introduce non-Cartesian coordinates (e.g., in a 3-D Cartesian $\vec{W} = x^\beta \hat{i} + y^\beta \hat{j} + z^\beta \hat{k}$ system, one can define $r^{2\beta} = \vec{W} \bullet \vec{W}$, along with angular variables to obtain a "spherically symmetric diffusion operator"). As in quantum mechanics, these should be derived by a coordinate transformation of the Cartesian-like Laplacians.

Finally, we are currently exploring the more general and difficult case of non-linear diffusion equations using our generalized Fourier transform methods.

**Acknowledgements**
This research was supported by the R. A. Welch Foundation under Grant E-0608. The author, DJK, gratefully acknowledges J. R. Klauder and G. H. Gunaratne for helpful conversations. The authors declare that no conflict of interests exists with the results and conclusions presented in this paper. Publication ethics have been observed.

**APPENDIX A**

In this Appendix we show the simple result that for any pair of non-commuting, canonically conjugate Cartesian-like variables, such as $W$, $P_W$, their quantum operators generate a Fourier-like transform for which the minimizing quantum state is a Gaussian that is invariant under the transform. Since the variables are assumed to be Cartesian-like and canonically conjugate, they result in the commutator given in Eq. (17). The minimizing condition is

$$\hat{W}\left|\psi_{\min}\right\rangle = -i\hat{P}_W\left|\psi_{\min}\right\rangle . \quad (A.1)$$

Inserting the resolutions of the identity in the W- and $P_W$-representations, one obtains

$$\int_{-\infty}^{\infty} dW\, W\left|W\right\rangle \psi(W) = \int_{-\infty}^{\infty} dK\, K\left|K\right\rangle \hat{\psi}(K) . \quad (A.2)$$

If we project this onto an eigenket of $\hat{W}$, we obtain

$$W\psi_{\min}(W) = -i\int_{-\infty}^{\infty} dK\, K\left\langle W|K\right\rangle \hat{\psi}_{\min}(K) . \quad (A.3)$$

Projecting Eq. (A.1) onto $\left|K\right\rangle$ results in the analogous equation

$$-iK\hat{\psi}(K) = \int_{-\infty}^{\infty} dW\, W\left\langle K|W\right\rangle \psi(W) . \quad (A.4)$$

By Eqs. (21) - (22), Eqs. (A.3) and (A.4) result in the W-wavefunction and the K-wavefunction being Gaussians of the exact same form.

24